# Heuristic arguments in favor of the view that the forces which form the bound systems of elementary particles propagate instantaneously

Ivan Aničin
*Institute of Physics, Belgrade, Serbia*

**Abstract**

We make use of the well-known properties of spontaneous decays of stationary states of bound systems of elementary particles to support the view that the forces which form these systems propagate instantaneously at the moment of their decay.

## Introduction

The question whether the forces between distant bodies propagate instantly or at a finite speed was a matter of dispute ever since the times of Newton [1]. There exist a number of comprehensive reviews on the subject [2, 3]. After the advent of relativity it has become generally agreed that neither a body nor an interaction can propagate at the speed greater than the speed of light, except perhaps for the problematic tachyons [4]. As it turned out, this actually holds for real particles only, which lie "on the mass shell". Quantum physics opened new issues. Nowadays it is considered, as is seemingly confirmed by recent measurements [5], that the Einsteinian "spooky action-at-a-distance", which connects the entangled pairs of real quantum particles and manifest quantum non-locality, does propagate at speeds that are very much greater than the speed of light, and possibly infinite. This is tantamount to the view that the collapse of a composite wave-function takes place instantaneously. Also, quantum field theories as an essential ingredient incorporate the unobservable space-like virtual particles, which violate the relativistic energy-momentum relations and are "off the mass shell", which may in principle propagate at arbitrary speed (for recent results see for instance [6]). Here, we present some heuristic arguments which support the view that static forces that bind the systems of elementary particles are necessarily transmitted by such particles.

Only two interactions are involved in binding the particles into complex systems – the electromagnetic and the strong, or rather the nuclear force, since we shall not consider the nucleons as composite particles. Weak interaction, although on the same footing with these, is not a binding force, and is not directly involved in our argumentation. Gravitation is, at least currently, conceptually completely outside of this scheme, and it may not be claimed that what we discuss here does in any way pertain to it.

## Weird properties of decays

We limit our discussion to the processes of spontaneous decays of stationary states of bound systems of particles, mostly those of nuclei and atoms. The decays result from the workings of the interactions within the system – from the workings of electromagnetic interaction (like in photon emission), strong interaction (like in alpha decay) and weak interaction (like in beta decays). Common to all these decays, however, is the peculiar nature of the decay process and it is this peculiarity, as we shall argue, which suggests that the interactions that bind the decaying systems propagate instantaneously throughout the system.

Stationary states of a quantum system of particles are built by the actions of ever-present static forces, the eternal sources of which are the particles themselves, and while the system resides in any of its stationary states the speed at which these forces propagate is meaningless. Some insight into the speed with which the interactions propagate may be gained only through the properties of the dynamic processes that go on within the system while it spontaneously changes its state. In what follows we shall, following the nuclear tradition, all such spontaneous transitions from stationary states of higher energy to the states of lower energy, call the decays. It is only in decays that we actually witness the formation of the systems of particles in a given stationary state, which is the final state of the transition.

Key to our argument is the incomprehensible property of quantum decays (or the once infamous "quantum jumps") that they do not take time – transitions between the stationary states of bound systems take place in a single point of space-time, or in the language of field theory diagrams, in a defined vertex, where all the changing quantities that describe the involved systems are conserved (if some quantity is not conserved it is again submitted to a definite non-conservation principle). Probability for such an instantaneous and sudden event, which is calculable by methods of perturbation theories, is always one and the same, as are all the static properties of stationary states. This results in the exponential distribution for survival times of a particular state, with the constant which is the sum of probabilities for all independent decay modes, which may be many. Applied to an ensemble of identical systems this leads to the law of radioactive decay, as it is called when applied to the phenomenon of radioactivity.

**Decay is sort of a timeless collapse into a particular final state**

It is perhaps the greatest riddle of them all that although any single system may decay by only one of the allowed decay modes, the energy width of the decaying state is always determined by the sum of probabilities for all allowed modes, realized or not (the width of the final state, if not stable, contributes to energy width of decay products as well, although none of its decay modes is realized yet). This suggests that in the given stationary state the system at all times possesses the necessary information about all possible final states of its decay and that at the instant of the decay it collapses to the particular one of them, as guided by the corresponding partial probability. The instant of the decay is, however, governed by the sum of probabilities for all possible decay modes. The collapse, then, is an instantaneous event that itself does not take time. Quantum world indeed appears so very far from classical mechanistic and causal considerations.

**Final state systems recoil as ideally rigid bodies**

Another argument, seemingly not connected to our previous discussion, comes from the often without saying assumed fact that in decays the system in its final state always recoils as an ideally rigid body – it exhibits a single and well defined mass, which has already taken account of all the rearrangements and changes in binding energies that might have taken place in the whole of the system, however complex these might have been. And all this in no time! Existence of ideally rigid bodies is banned by relativity, since they require instantaneous spread of information throughout the body about its overall mass. That quantum systems behave that way is another hint towards the instantaneous nature of the interactions that govern both their structure and dynamics. That the masses of resulting systems are instantaneously adjusted to satisfy the conservation laws, implies that the mass-determining interactions, comprising the transformation of complex mass defects into kinetic energies of the final state systems, are not only ever-present, as the static situations imply, but that also propagate instantaneously in a dynamic situation like the decay.

Let us take simple alpha decay as an example. The two fragments of the initial nucleus recoil with their masses completely defined at the instant of the decay, in accord with all conservation laws. Since, as in any decay, the available energy (the Q-value) depends on both the initial and final masses this implies that at the instant of decay all of them coexist in the nucleus. As the process is necessarily realized via tunneling, which involves the transmission of an evanescent wave, what has again been shown not to take any measurable time [7], everything seems consistent.

Next we take the simplest example from the realm of the processes which are usually termed atomic, namely the deexcitation of the hydrogen atom. If, for instance, there occurs a transition which results in the emission of a visible photon with an energy of the order of 2 eV then, if only the electron participated in the emission, the energy of the recoil would be of the order of the quite noticeable $2^2/2 \times 511000$, which is obviously not the case. That would, among other things, prevent resonant absorption, since natural widths of atomic states are significantly smaller than this value. Since the whole atom participates, as we usually assume without saying, the recoil of $\sim 2^2/2 \times 10^9$ does not measurably influence the photon energy. If emission would be a chain two-step process, and the electron recoil was transferred adiabatically to the proton, with the corresponding delay, the photon would not have the information of the nucleus' presence at the instant of emission, and the comparatively large electron recoil would measurably decrease the photon energy, disabling the resonant absorption.

The most conspicuous example of similar argument is perhaps provided by the Moessbauer effect. Emission of a gamma-ray from a nucleus whose atom is free results in the recoil of the atom with an energy which is small, but is sufficiently large as compared to natural widths of the involved states, so as to prevent resonant absorption in other identical atoms. If the atom is embedded in a crystal lattice then there is a probability (the so-called recoilless fraction) that the recoil would not excite the lattice, and the crystal would recoil as a whole. This leaves the energy of the gamma-ray virtually unaffected, what results in resonant absorption in similarly bound atoms. We see that now the whole macroscopic body recoils as a single ideally rigid body with a macroscopic mass, which is well defined at the instant of the gamma decay that took place in one of its atoms. Even in this case we are compelled to conclude that the mass-defining interaction, which is here a complex interaction of electromagnetic origin, propagates instantaneously throughout the whole macroscopic system.

**Decays are neither purely atomic nor purely nuclear events**

Atomic electrons are found in the near field zone of the nucleus, and, like in nuclear magnetic resonance, where the system of nuclear spins feeds and drains energy in and out of the macroscopic oscillator by the exchange of virtual photons, the nucleus and atomic electrons form the single bound system. This is perhaps best illustrated by the processes of internal conversion, where atomic electrons drain the energy out of a nuclear excitation, and vice versa, in NEET processes (Nuclear Excitation by Electron Transitions), where atomic electrons feed the energy of their excitation into the nucleus. This author had the opportunity to discuss these matters with Mel Freedman of Argonne, who in the early fifties of the last century was the first to resolve the dilemma that existed at that time about the detailed distribution of energies between different participants in the final state of the beta-decaying atom. Ever since, Freedman strongly advocated the view, which is nowadays generally shared, that the so called nuclear processes always involve whole atoms, and that in the wrongly termed "nuclear decays" both the system of the nucleus and that of atomic electrons get transformed in a complex way, in a single step process [8]. In his own words: "The nuclear and atomic transitions are simultaneous processes in the composite nucleus-atom system". This view alone corroborates our conjectures about the instantaneous actions of the forces that manifest themselves in the decays of bound systems of particles.

The mentioned episode with Mel Freedman is instructive in this respect. In the early fifties of the last century it was not clear whether in beta decay the atom recoils in the state which preceded the decay and that it only then in a second stage gets adiabatically excited due to the change of charge in its nucleus, or that it recoils in the excited state simultaneously with the emission of the nuclear lepton pair, in a single step process. Being able to demonstrate that the beta spectrum of Pu-241 begins at zero energy, meaning that the difference in total atomic binding energies in the initial and final states end up in the energy of the emitted lepton pair, the authors [9] proved that the beta-decay is a single step process, which embraces simultaneous changes in both the nuclear and electron structures of the decaying atom. This knowledge later greatly helped Freedman to correctly elucidate the mechanisms of initial processes in recoil chemistry.

Another striking example of the unity of nuclear and atomic systems is the exotic process of the bound beta decay, which Freedman also tackled successfully [10]. The case in question, among the few where the process is at all possible, was the case of the thallium-205 atom, which is, when the atom is complete, a stable one, with a negative Q-value for any possible mode of decay. If it is, however, gradually deprived of its electrons and if the nucleus finally remains bare, the Q-value suddenly becomes positive for a peculiar form of the nuclear beta-minus decay, in which the beta electron ends up in the bound state of its own atom. Properties of the otherwise stable nuclear stationary state thus completely change at the expense of binding energies of atomic electrons. The resulting atom is now the hydrogen-like lead-205, which is in this form stable, while after completing its electron structure the neutral lead-205 then beta-decays (by electron capture) back to thallium-205!

**Structureless particles follow the same decay rules**

One final argument, which could as well have been the first one, comes from the world of truly fundamental and, as is presently believed, structureless fundamental particles, which share the same decay rules with the complex systems that we have just described. Good example is the heaviest of leptons, the tau. It has many decay modes but a single energy width and a unique lifetime. As there is no dynamics in the processes that go on in a structureless entity, complex entities that follow the same rules should not have any dynamics as well.

**Conclusions**

We believe that the few selected examples of decays that we briefly analyzed served our aim, which was to demonstrate that the forces, which govern the structure of the systems and their decays, necessarily propagate instantaneously throughout the whole system, however complex and heterogeneous it might be. The discussed properties naturally extend to molecules, where, for instance, nuclear properties determine the character of molecular rotational spectra, but this would tell us nothing new in principle and we do not pursue the issue further.

A word about terminology is perhaps in order in the end. From the early days of quantum physics it was clear that our language, which was developed as the means to describe the causal phenomena in the world of our senses, is ill suited to description of the submicroscopic world. Starting with the title to this paper, and throughout, we suffer from the same inadequacy. The phrase that "forces propagate instantaneously" is a misnomer, for there is no such thing as instantaneous propagation – what is instantaneous does not propagate, for it is at that very instant everywhere. The statement that "decays do not take time", is perhaps only a slightly better way to express the nature of the forces that govern what goes on in the system when it comes to changing its state. Calling this "a process", what we also use, again does not seem adequate, for the process usually denotes a sequence of states, what necessarily takes some finite time, what the decays of stationary states of quantum systems do not do. Due to this, all attempts to dynamically describe the decays themselves should end up in divergences. In short, our thesis may be, in the operational spirit, best summed up as follows: In bound systems of particles time manifests itself as an observable towards the outside world only via finite lifetimes of their excited states, or via the corresponding decay probabilities, while the goings-on in the system are timeless – stable states live eternally, while changes of unstable states do not take time. Time in microscopic systems thus appears to be of probabilistic character only, while its dynamic definition seems void of operational meaning. This looks like another manifestation of statistical determinism – it is not only that the proverbial God throws dice when He is to decide the outcome of a given decay, but one even cannot know *when* He will throw one. Probability itself again appears as a genuine physical quantity. It is only thanks to the law of great numbers that the behavior of bound systems of particles may define the smooth flow of time. Whether or not all this may be said to imply the instantaneous propagation of the mass-defining interactions within the systems remains a matter of linguistics.

Discussions with Prof. Dj. Krmpotić are greatly acknowledged.

**References**

[1] Steffen Ducheyne
*Newton on Action at a Distance*,
Journal of the History of Philosophy, Volume **52**, Number 4, October 2014, pp. 675-701

[2] Mary Hesse
*Forces and Fields – The Concept of Action at a Distance in the History of Physics*,
Nelson, London and New York, 1961

[3] F.Hoyle and J.V.Narlikar,
*Action at a distance in Physics and Cosmology*
Freeman, San Francisco 1974

[4] Moses Fayngold
*Special Relativity and Motions Faster than Light*
Wiley-VCH Verlag GmbH,
Weinheim, 2002

[5] Juan Yin et al,
*Bounding the speed of 'spooky action at a distance'*

[6] J.H.Field
*Comment on 'Measuring propagation speed of Coulomb fields'*

[7] Horst Aichmann and Günter Nimtz,
*The Superluminal Tunneling Story*,

[8] M S Freedman
*Atomic Structure Effects in Nuclear Events*
Annual Review of Nuclear Science Vol. **24**: 209-248 (1974)

[9] Melvin S. Freedman, F. Wagner, Jr., and D. W. Engelkemeir
*The Beta-Spectra of Pu239, Pu240, and Pu241*
Phys. Rev. **88**, 1155 (1952)

[10] Melvin S. Freedman,
*Measurement of low energy neutrino absorption probability in Thallium 205,*
Nucl. Instr. and Meth. in Phys. Res. **A271**(1988)267